# Giant excitonic effects in bulk vacancy-ordered double perovskites

Fan Zhang[1], Weiwei Gao[1*], Greis J. Cruz[3], Yi-yang Sun[2], Peihong Zhang[3*], Jijun Zhao[1]

1. Key Laboratory of Material Modification by Laser, Ion and Electron Beams (Dalian University of Technology), Ministry of Education, Dalian 116024, China
2. State Key Laboratory of High Performance Ceramics and Superfine Microstructure, Shanghai Institute of Ceramics, Chinese Academy of Sciences, Shanghai 201899, China
3. Department of Physics, University at Buffalo, State University of New York, Buffalo, New York 14260, USA

**ABSTRACT**

Using first-principles GW plus Bethe-Salpeter equation calculations, we identify anomalously strong excitonic effects in several vacancy-ordered double perovskites $Cs_2MX_6$ (M = Ti, Zr; X = I, Br). Giant exciton binding energies about 1 eV are found in these moderate-gap, inorganic bulk semiconductors, pushing the limit of our understanding of electron-hole (*e-h*) interaction and exciton formation in solids. Not only are the exciton binding energies extremely large compared with any other moderate-gap bulk semiconductors, but they are also larger than typical 2D semiconductors with comparable quasiparticle gaps. Our calculated lowest bright exciton energy agrees well with the experimental optical band gap. The low-energy excitons closely resemble the Frenkel excitons in molecular crystals, as they are highly localized in a single $[MX_6]^{2-}$ octahedron and extended in the reciprocal space. The weak dielectric screening effects and the nearly flat frontier electronic bands, which are derived from the weakly bonded $[MX_6]^{2-}$ units, together explain the significant excitonic effects. Spin-orbit coupling effects play a crucial role in red-shifting the lowest bright exciton by mixing up spin-singlet and spin-triplet excitons, while exciton-phonon coupling effects have minor impacts on the strong exciton binding energies.

Exciton is a composite boson consisting of a correlated electron-hole *(e-h)* pair bounded by screened Coulomb interaction. The attractive Coulomb interaction between electrons and holes creates low-lying excitons with energy $E_{ex}$ below the quasiparticle band gap $E_g^{QP}$ (i.e., fundamental gap). The binding energy of the lowest-energy exciton is typically defined as the offset between the fundamental gap and the exciton energy: $E_b^{ex} = E_g^{QP} - E_{ex}$. Systems with large exciton binding energies are important for exploring light-matter interactions as well as for developing exciton-based optoelectronic devices operating at room temperature [1–8]. Moreover, they are potential hosts for composite quasiparticles like bi-excitons [9,10], trions [11,12], and exciton-polaritons [13,14], which have been extensively studied during the last decade. Excitonic effects also profoundly impact the performance of optoelectronic materials such as quasi-two-dimensional (2D) hybrid perovskites [15–17].

Well-known examples of materials with significant excitonic effects include low-dimensional materials (such as semiconductor nanoclusters, carbon nanotubes, and 2D semiconductors) [18–21], molecular crystals [22,23], and alkaline halides [24,25]. The latter two are bulk materials that can host strongly bounded Frenkel or charge-transfer excitons. On the other hand, most all-inorganic semiconductors with moderate fundamental band gaps ($E_g < 4$ eV) exhibit Wannier-Mott excitons, which have small exciton binding energies of a few meV to 100 meV [26–28]. Notably, cuprous chloride and cuprous oxide, which have been extensively studied for realizing Bose-Einstein condensation of excitons, possess large exciton binding energy of nearly 200 meV and demonstrate prominent Rydberg exciton series [29–31]. Other inorganic semiconductors with large exciton binding energies include delafossite compounds ($E_b^{ex} = 0.3 \sim 0.5$ eV) [32,33] and quasi-2D all-inorganic perovskites [34–36]; all of them have quasi-layered structures. However, compared to typical one-dimensional (1D) or 2D semiconductors, exciton binding energies of bulk inorganic materials with moderate gaps are still much weaker.

Owing to the rich choices of chemical composition, tunable optical band gap between 1 to 3 eV, non-toxicity, and earth abundance of the elements, the synthetic vacancy-ordered double perovskites (VODP) have been proposed as promising materials for photovoltaic applications [37–39]. Using density functional theory (DFT) [40] and many-body perturbation theory (MBPT) [41,42], in this Letter we investigate the quasiparticle and excitonic properties of a group of vacancy-ordered double perovskites (VODP) materials, including $Cs_2TiX_6$ and $Cs_2ZrX_6$ (X = I, Br). Recent studies showed a surprisingly large discrepancy (over 1 eV) between the calculated fundamental band gaps of $Cs_2TiI_6$ and $Cs_2ZrI_6$ and the measured values [37,43]. The overestimation of fundamental band gaps was ascribed to improper treatment of localized Ti 3$d$ orbitals within the GW approximation [37]. Our combined GW and Bethe-Salpeter equation (BSE) [44–46] calculations, however, reveal an extremely strong excitonic effect in these materials with exciton binding energies far exceeding those in other moderate-gap bulk semiconductors, thus bringing the theoretical optical gaps in line with experimental results. The discovery of 3D all-inorganic semiconductors with moderate gaps and anomalously strong excitonic effects not only broadens our understanding on excitonic effects in solids but also provides clues in searching for bulk systems with large exciton binding energies.

The crystal structure of $A_2MX_6$ (A = Cs, M = Ti, Zr, and X = I, Br) VODP can be obtained by removing every other M-site cation from $AMX_3$ halide perovskites [38,43], as shown in Fig. 1(a). The $[MX_6]^{2-}$ octahedral clusters in VODP $A_2MX_6$ do not form covalent bonds with one another because they are separated by the evenly distributed M-site vacancies. Alternatively, one may consider $Cs_2MX_6$ as an $A_2B$ antifluorite structure, where Cs ions occupy the A sites and $[MX_6]^{2-}$ octahedra occupy the B sites.

The band structures of these four VODPs are quite similar, and replacing Ti (I) with Zr (Br) in $Cs_2MX_6$ increases the fundamental band gap [37]. Both $Cs_2TiX_6$ and $Cs_2ZrX_6$ have an indirect band gap between the valence band maximum (VBM) at $\Gamma$ and the conduction band minimum (CBM) at $X$. The low-energy conduction bands are mainly derived from the M-site $d$ states, whereas the highest valence bands are mostly contributed by the halogen element X. Under the influence of the octahedral crystal field, the $d$-derived conduction bands split into a doublet and a triplet group; the low energy triplet bands are fairly flat with an extremely narrow band width (0.29 eV, 0.25 eV, 0.44 eV, and 0.37 eV for $Cs_2TiI_6$, $Cs_2TiBr_6$, $Cs_2ZrI_6$, and $Cs_2ZrBr_6$ respectively). The top valence band is virtually dispersionless from $\Gamma$ to $X$. The direct gap of $Cs_2TiI_6$ at $\Gamma$ is only 40 meV larger than the indirect gap between $\Gamma$ and $X$, resulting in a "quasi-direct-gap" band structure.

To better illustrate the roles of $Cs^+$ and $[MX_6]^{2-}$ on the frontier states, we compare the DFT band structure of $Cs_2TiI_6$ with those of two fictitious systems, namely, a periodic system of $[TiI_6]^{2-}$ without the $Cs^+$ ions, and an isolated $[TiI_6]^{2-}$ octahedron, as shown in Fig. 1(b). The DFT band structures are calculated using the Perdew–Burke-Ernzerhof (PBE) functional [47] without including spin-orbit coupling (SOC). The nearly identical frontier band structures of $Cs_2TiI_6$ and $[TiI_6]^{2-}$ indicate that $Cs^+$ ions merely function as electron donors and spacers, but do not significantly affect the band dispersion of the system. The role of $Cs^+$ is similar to that of $CH_3NH_3^+$ in $CH_3NH_3PbI_3$ [48]. The highest-occupied molecular orbital (HOMO) of $[TiI_6]^{2-}$ is mostly composed of the iodine $5p$ orbitals, while titanium $3d$ orbitals dominate the lowest-unoccupied molecular orbital (LUMO). These $[TiI_6]^{2-}$ derived LUMO and HOMO states interact to form the lowest conduction bands and highest valence bands of $Cs_2TiI_6$, respectively. The narrow widths of the frontier energy bands in $Cs_2TiI_6$ further suggest that $[TiI_6]^{2-}$ behaves like a polyatomic superanion [49], which interacts weakly with each other. In this manner, $Cs_2TiI_6$ crystal can be understood as a three-dimensional (3D) assembly of $[TiI_6]^{2-}$ units and $Cs^+$ spacers. As we will discuss later, such a unique structure is responsible for extremely strong excitonic effects in VODPs.

The quasiparticle (QP) band structure of $Cs_2TiI_6$ calculated with the $G_0W_0$@PBE method including SOC effect is shown in Fig. 1(c). The indirect QP band gap of $Cs_2TiI_6$ is 1.79 eV, which is slightly smaller than the direct band gap at $\Gamma$ (1.83 eV). The calculated QP band gap within the $G_0W_0$ approximation is significantly larger than the measured optical gap of 1.02 eV [43]. It is noteworthy that our calculated QP gap is smaller than that reported in a previous work [37] by 0.54 eV. This discrepancy mainly stems from the use of different cut-off parameters and treatment of frequency-dependent dielectric function [50–53]. Indeed, a small cut-off of dielectric matrices used in GW calculations can lead to an overestimation of quasiparticle band gap by about 0.5 eV (more details can be found in Fig. S1 and the corresponding discussions in Supplemental Material). We also compare the QP band gaps of all four VODPs in Table 1 and the DFT-PBE band structures in Fig. S3 of Supplemental Material.

A comparison of the band structures calculated with and without SOC (Fig. S3 of Supplemental Material) reveals strong relativistic effects on the top valence bands, and the gap at the $\Gamma$ point of $Cs_2TiI_6$ is reduced by about 130 meV due to the SOC effects. Fig. 1(d) highlights the impacts of the SOC effects on the frontier electronic states at the $\Gamma$ and $X$ points of $Cs_2TiI_6$. If the SOC effects are neglected, the lowest-energy direct transition at the $\Gamma$ and $X$ points, which are $\Gamma_{15'} \rightarrow \Gamma_{25'}$ and $X_4 \rightarrow X_3$, respectively, are dipole forbidden, while the second lowest-energy direct transition at the $\Gamma$ point ($\Gamma_{15} \rightarrow \Gamma_{25'}$), is dipole allowed [54]. After the SOC effects are considered, the triply degenerate states (i.e., $\Gamma_{15}$, $\Gamma_{15'}$, and $\Gamma_{25'}$) split into quadruple and double degenerate states (including Kramer's degeneracy). Considering the symmetry properties of the SOC-split states, the dipole-allowed optical transitions are indicated by green arrows in Fig. 1(d). More details on the optical selection rules can be found in Supplemental Material. Due to the SOC splitting, the energy of the lowest dipole-allowed transition $\Gamma_8^- \rightarrow \Gamma_8^+$ is smaller than that of $\Gamma_{15} \rightarrow \Gamma_{25'}$ by about 0.31 eV. In other words, the relativistic effects tend to red-shift the absorption energy within the single-particle-transition picture. In the following, we show that this trend also holds even when the excitonic effects are considered.

Fig. 2(a) shows the imaginary part of the frequency-dependent dielectric constant $\varepsilon_2(E)$ of $Cs_2TiI_6$, calculated using the GW+BSE approach and independent-particle approximation with inclusion of SOC effects. Our GW+BSE calculations reveal several prominent excitonic absorption peaks far below the quasiparticle band gap. In fact, the lowest-energy exciton locates at 0.85 eV, indicating a giant binding energy of nearly 1 eV, which is an important character of Frenkel excitons. This is a dark exciton with a negligible optical dipole element, while the lowest-energy bright exciton locates at 1.04 eV, agreeing well with the experimental optical gap of 1.02 eV [43]. We have carefully checked the convergence of the calculated exciton binding energies and absorption spectra and found that the results converge quickly with respect to the density of *k*-point sampling (as shown in Table S2 and Fig. S2 of Supplementary Material). In contrast, typical Wannier excitons in bulk and 2D systems often require extremely dense *k*-grids to achieve proper convergence of the calculated exciton binding energies [55,56]. As we will show later, the *e-h* amplitudes of the low energy excitons spread across a large part of the Brillouin zone (BZ), which partially explains the rapid convergence behavior of exciton binding energy.

The strong SOC effects on the calculated band structure, especially on the top valence states as shown in Fig. 1(c), result in significant changes in the calculated excitonic structure and optical absorption. To uncover the effects of SOC on the low-energy excitons, we compare the GW+BSE results calculated with and without the SOC effects. When the SOC effects are neglected, an excitonic state can be either spin-0 (singlet) or spin-1 (triplet). The triplet excitons are always dark since spin is conserved in optical dipole transitions if the SOC effects are neglected. The excitation energies of the singlet and triplet excitons for $Cs_2TiI_6$ are shown in Fig. 2(b), in which each vertical line corresponds to an excitonic state; these lines are colored according to their brightness, i.e., the square of the dipole matrix elements. The lowest-energy triplet exciton is about 0.1 eV lower than the lowest-energy singlet one, which is also a dark exciton due to the orbital symmetry as discussed earlier. The lowest-energy bright exciton is located at around 1.34 eV and can be attributed to the $\Gamma_{15} \to \Gamma_{25'}$ transitions. When the SOC effects are considered, spin is no longer conserved, and the excitonic states are a mixture of spin-singlets and spin-triplets [45]. As displayed in Fig. 2(c), the lowest-energy exciton, which mainly derives from the low-energy spin-triplet states, is dark. Moreover, the SOC effects significantly mix the singlet and triplet excitons and red-shift the absorption edge from 1.34 eV to 1.04 eV, bringing theory in better agreement with experiment [43].

The calculated energies of the lowest-energy bright excitons for the other three VODPs ($Cs_2TiBr_6$, $Cs_2ZrI_6$ and $Cs_2ZrBr_6$) also agree well with the experimental optical band gaps, as summarized in Table 1. The imaginary parts of the dielectric functions of them are compared in Fig. S4-S6 of Supplemental Material. Similar to the case of $Cs_2TiI_6$, the lowest-energy excitons of these three VODPs are also dark, and the corresponding exciton binding energies are extremely large. The exciton binding energy of $Cs_2TiI_6$ ($E_b^{ex}$ = 0.98 eV) is nearly twice that of monolayer 2H-MoSe$_2$ ($E_b^{ex}$ = 0.55 eV) [5], which has a comparable quasiparticle band gap ($E_g$ = 2.1 eV) as $Cs_2TiI_6$. This finding is rather surprising, as it is well known that the excitonic effects in bulk semiconductors are often much weaker than those in 2D semiconductors [57–59] owing to the stronger screening effect. Perhaps more striking is the binding energy of the lowest dark exciton in $Cs_2ZrBr_6$, which is as large as 1.69 eV. The abnormally large exciton binding energies in these moderate-gap VODPs deserve closer scrutiny.

To gain a deeper insight into the low-energy excitons in VODP, we examine their wave functions in both reciprocal and real spaces. Within the Tamm-Dancoff approximation [45,60], an exciton wave function can be expanded as a linear combination of products of electron wave function $\psi_{ck}(r_e)$ and hole wave function $\psi_{vk}(r_h)$:

$$\psi^S(r_e, r_h) = \sum_{k,c,v} A^S_{vck}\psi_{ck}(r_e)\psi^*_{vk}(r_h), \quad (1)$$

where $S$ is the index of an excitonic state and $A^S_{vck}$ are the coupling coefficients between electron and hole states. We have analyzed the BZ distribution of the lowest-energy dark and bright excitons in VODPs by defining a $k$-dependent $e$-$h$ amplitude $A^S(k) = \sum_{vc}|A^S_{vck}|^2$, as shown in Fig. 3(a). The radii of the circles are proportional to $A^S(k)$. In contrast to halide perovskites, in which the Wannier-Mott excitons are highly localized in small regions of the BZ [61–63], the low-energy excitons in $Cs_2TiI_6$ are far more extended in the BZ. As can be seen from Fig. 3(a), the direct transitions at Γ point contribute the most to the lowest dark and bright excitons, while contributions from direct transitions at the BZ boundary are also considerably large. To see the differences between conventional Wannier excitions and the Frenkel excitons in VODPs, we compare the $k$-space wave function $\Psi_{ex}(k)$ with a hydrogenic model [64,65] $\Psi_{hy}(k) = (2a_0)^{\frac{3}{2}}/\pi(1 + a_0^2 k^2)^2$. The modular square of wave function $|\Psi_{ex}(k)|^2$ is proportional to $A^S(k)$ and normalized such that $\int |\Psi_{ex}(k)|^2 d^3k = \sum_k A^S(k) = 1$. As shown in the inset of Fig. 3 (a), the calculated $|\Psi_{ex}(\mathbf{k})|^2$ of the lowest-energy exciton agrees reasonably with $|\Psi_{hy}(\mathbf{k})|^2$ at small $|\mathbf{k}|$, but deviates from the hydrogenic model for large $|\mathbf{k}|$. Here the fitted hydrogenic model $\Psi_{hy}(\mathbf{k})$ has a small fitted Bohr radius $a_0 = 3.7$ Bohr, suggesting a small spatial extent of the exciton.

To visualize the real-space distribution of low-energy excitons, we fix the hole position at an iodine atom, and plot the electron distribution of the lowest-energy dark and bright excitons for Cs$_2$TiI$_6$ in Fig. 3(b), which shows that the electron is mainly localized within one [TiI$_6$]$^{2-}$ octahedron. Such a localized real-space distribution is in line with our previous analysis of the electronic structure. As we have discussed earlier, Cs$_2$MX$_6$ can be considered as a 3D assembly of weakly coupled [MX$_6$]$^{2-}$ clusters, resulting in the formation of Frenkel excitons like those in molecule solids [66]. In addition, the relatively weak local dielectric screening (as demonstrated by the high-frequency dielectric constants in Table 1) also contributes to an overall strong *e-h* interaction, thus the large exciton binding energies.

As recently shown by Marina *et al.* [67], exciton-phonon coupling can considerably renormalize the exciton binding energies in ionic materials like GaN and CsPbI$_3$. One may question whether the exciton-phonon coupling can significantly affect the exciton binding energies of VODPs. We estimate the correction ($\Delta E_b$) to exciton binding energy due to phonon-screening effects using a simplified formula proposed by Marina *et al.* [66]:

$$\Delta E_b = -2E_b \frac{\omega_{LO}}{\omega_{LO} + E_b}\left(1 - \frac{\varepsilon_\infty}{\varepsilon_0}\right), \quad (2)$$

where the high-frequency dielectric constant $\epsilon_\infty$ is calculated within the random-phase approximation, $\epsilon_0$ is the static dielectric constant, and $\omega_{LO}$ is the frequency of the dominant longitudinal optical (LO) phonon mode, which is usually the highest LO mode. We show in Table 1 the estimated $\Delta E_b$, which are less than 50 meV for all four systems studied. Therefore, while these corrections are not negligible, they do not significantly affect our conclusion that the exciton binding energies in VODPs are exceedingly large compared to any other known moderate-gap 3D semiconductors.

In summary, we have predicted giant exciton binding energies ranging from 0.95

to 1.65 eV (after corrected for the electron-phonon renormalization effects) in moderate-gap bulk VODP materials $A_2MX_6$ (A = Cs; M = Ti, Zr; X = I, Br). The exciton binding energies in these systems are one order-of-magnitude larger than typical inorganic semiconductors with comparable quasiparticle band gaps; they are even larger than those of monolayer transition metal dichalcogenides with similar fundamental band gaps. The lowest-energy excitons are dark, and the predicted absorption edges agree well with experiment, resolving an outstanding puzzle that the calculated (quasiparticle) band gaps seem to be much larger than the measured (optical) gaps. Spin-orbit coupling effects play an important role in mixing the spin-singlet and spin-triplet excitons, resulting in a red-shift to the absorption edges. According to the present findings, we envision the future discovery of more inorganic semiconductors with enormous exciton binding energies. Such semiconductors with giant excitonic effects possess two potential characteristics. First, its crystalline structure can be viewed as an assembly of weakly coupled low-dimensional units, such as superatomic clusters [49] or 1D wires. Second, its static dielectric constant should be small enough to ensure strong Coulomb interaction that binds *electron-hole* pairs. These materials, including bulk VODPs, provide potential platforms for exploring the properties and dynamics of Frenkel excitons, investigating more exotic composite quasiparticles such as bi-excitons and trions, and realizing exciton-based optoelectronics devices.

**ACKNOWLEDGMENTS**

This work is supported by the National Natural Science Foundation of China (12104080, 91961204) and XingLiaoYingCai Project of Liaoning province, China (XLYC1905014). Work at UB is supported by the National Science Foundation (DMREF-1626967). The authors acknowledge the computer resources provided by the Supercomputing Center of Dalian University of Technology and the Center for Computational Research, University at Buffalo, SUNY.

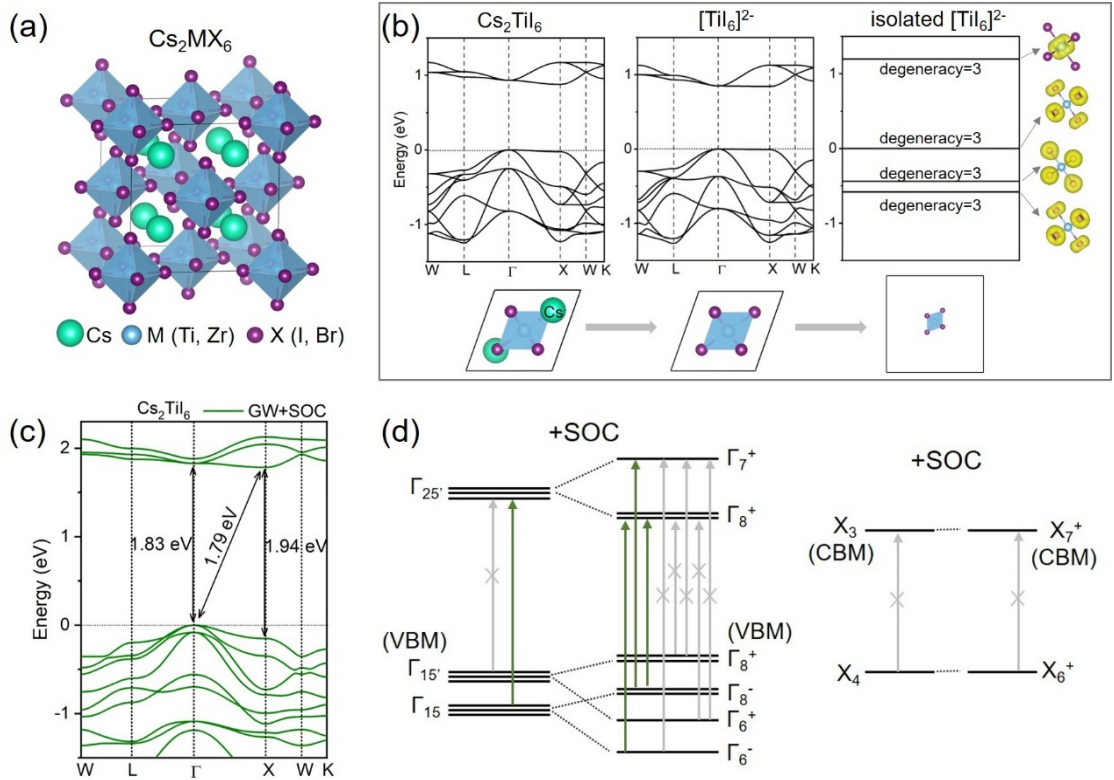

FIG. 1. (a) Crystal structures of $Cs_2MX_6$ showing the isolated octahedral units. (b) The DFT band structure of $Cs_2TiI_6$, $[TiI_6]^{2-}$ and isolated $[TiI_6]^{2-}$ with PBE functional. The Kohn-Sham wavefunction of isolated $[TiI_6]^{2-}$ is also plotted in the Figure. (c) GW quasiparticle band structure of $Cs_2TiI_6$ calculated with SOC, where W (0.5, 0.25, 0.75), L (0.5, 0.5, 0.5), Γ (0, 0, 0), X (0.5, 0.0, 0.5) and K (0.375, 0.375, 0.75) are high-symmetry k-points in the first Brillouin zone. (d) A schematic diagram showing the spin-orbit splitting and allowed optical transitions at Γ and X point of $Cs_2TiI_6$. For simplicity, the Kramer's degeneracy is not shown for the SOC-split spinor states.

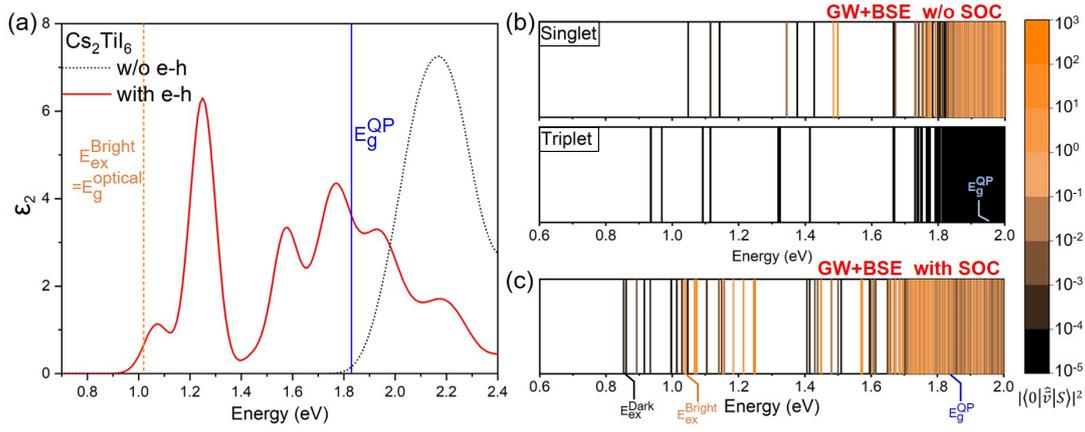

FIG. 2. (a) The imaginary part of dielectric function of $Cs_2TiI_6$, calculated without electron−hole interaction (black dashed line) and with electron-hole interaction (red solid line). A gaussian smearing of 0.05 eV is used for calculating the spectra. The lowest-energy bright exciton and fundamental band gap are marked in orange and blue lines, respectively. (b) Energies of singlet and triplet excitons calculated with a GW+BSE approach without SOC effects. (c) Energies of excitons calculated with a GW+BSE approach including SOC effects.

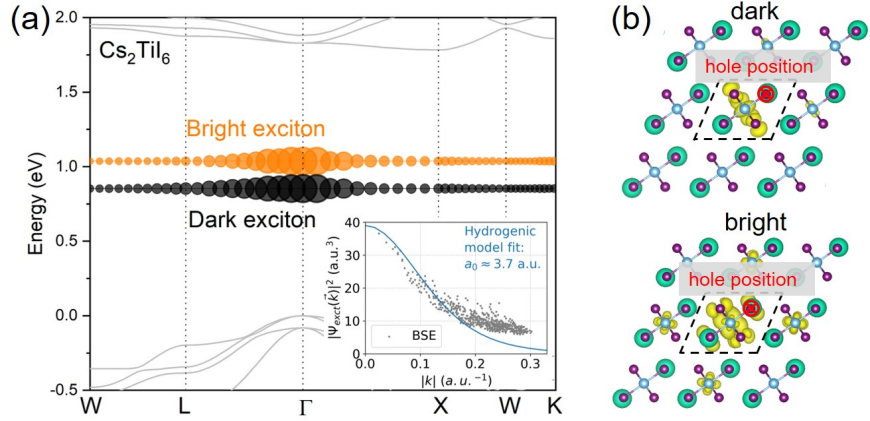

FIG. 3. (a) Reciprocal-space distribution of exciton wave functions $A^S(k)$ of Cs$_2$TiI$_6$. The orange and black circles represent the lowest bright and dark excitons, respectively. The inset is fitting the exciton wave function with hydrogenic model. (b) Real-space distribution of electron for the lowest-energy dark and bright excitons, respectively. The position of the hole is fixed at an iodine atom in Cs$_2$TiI$_6$. The position of hole is shown as a red cross here. The yellow surfaces represent the isosurfaces which include 80 percent of the electron charge density. The back dashed line shows the primitive cell.

Table 1. QP band gap, optical gap within experiment, energies and binding energies of lowest-energy dark and bright excitons calculated by GW+BSE approach, static ($\varepsilon_0$ from density functional perturbation theory; DFPT) and high-frequency ($\varepsilon_\infty$ from $G_0W_0$) dielectric constants, calculated longitudinal-optical (LO) phonon frequencies ($\omega_{LO}$), and the phonon screening correction of four VODPs. The direct band gap at $\Gamma$ and X, and the indirect band gap between X and $\Gamma$ are listed.

|  |  | $Cs_2TiI_6$ | $Cs_2TiBr_6$ | $Cs_2ZrI_6$ | $Cs_2ZrBr_6$ |
|---|---|---|---|---|---|
| Quasiparticle band gap (eV) | $\Gamma - \Gamma$ | 1.83 | 3.35 | 3.22 | 4.97 |
|  | X - X | 1.94 | 3.40 | 3.36 | 5.03 |
|  | X - $\Gamma$ | 1.79 | 3.31 | 3.20 | 4.95 |
| Optical gap (experiment) (eV) |  | 1.02 | 2.00 | - | 3.76 |
| $E_{ex}^{Bright}$ (eV) |  | 1.04 | 2.03 | 2.33 | 3.59 |
| $E_{ex}^{Dark}$ (eV) |  | 0.85 | 1.66 | 2.18 | 3.28 |
| $E_{b}^{Bright}$ (eV) |  | 0.79 | 1.32 | 0.89 | 1.38 |
| $E_{b}^{Dark}$ (eV) |  | 0.98 | 1.69 | 1.04 | 1.69 |
| $\varepsilon_\infty$ |  | 4.46 | 3.26 | 3.68 | 2.97 |
| $\varepsilon_0$ |  | 11.27 | 8.59 | 11.89 | 8.90 |
| $\omega_{LO}$ (meV) |  | 27.91 | 34.50 | 24.40 | 30.14 |
| $\Delta E_b$ (meV) |  | -32.8 | -42.0 | -32.9 | -39.5 |

# Supplemental Material

# Giant excitonic effects in bulk vacancy-ordered double perovskites

Fan Zhang[1], Weiwei Gao[1*], Greis J. Cruz[3], Yi-yang Sun[2], Peihong Zhang[3*], Jijun Zhao[1]

1. Key Laboratory of Material Modification by Laser, Ion and Electron Beams (Dalian University of Technology), Ministry of Education, Dalian 116024, China

2. State Key Laboratory of High Performance Ceramics and Superfine Microstructure, Shanghai Institute of Ceramics, Chinese Academy of Sciences, Shanghai 201899, China

3. Department of Physics, University at Buffalo, State University of New York, Buffalo, New York 14260, USA

* Corresponding Authors. Email: weiweigao@dlut.edu.cn; pzhang3@buffalo.edu

**Impacts of convergence and starting wave functions on $G_0W_0$+BSE results**

As demonstrated by many studies, calculations based on the GW approximation and Bethe-Salpeter equation (BSE) can depend sensitively on cut-off parameters and detailed implementations [1–4], such as the starting points of GW calculations. We investigate how the main cut-off parameters affect our computation results and conclusions presented in the main manuscript. To save computational costs, all our tests of the convergence and the investigation of various levels of approximations used in GW+BSE calculations are performed without the full-relativistic (i.e., spin-orbit coupling) effects. We assume the set of cut-off parameters that converge the GW+BSE calculation results without SOC effects can also converge those with SOC effects.

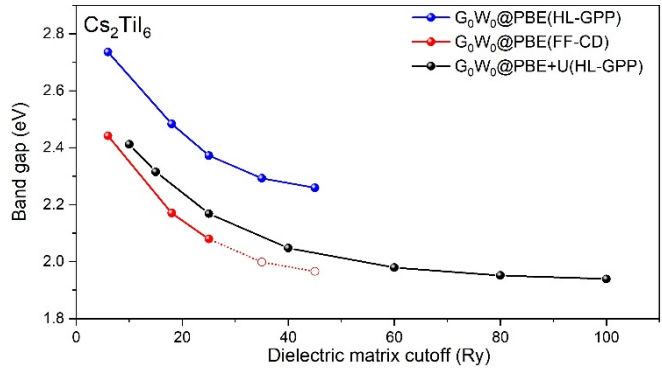

FIG. S1. The QP band gap of $Cs_2TiI_6$ as a function of the dielectric matrix cutoff by $G_0W_0$@PBE (HL-GPP, Hybertsen-Louie generalized plasmon-pole model), $G_0W_0$@PBE (FF-CD, full-frequency with Contour deformation approach) and $G_0W_0$@PBE+U (HL-GPP). Here the QP band gap represents the direct band gap at Γ.

Fig. S1 shows the quasiparticle (QP) band gap of $Cs_2TiI_6$ calculated with different kinetic-energy cutoffs of the dielectric matrices. As the dielectric matrix cutoff increases, the QP band gap at Γ decrease monotonically. If one uses a 6 Ry cutoff for dielectric matrices, the band gap would be overestimated by over 0.5 eV than the converged value. The band gap converges within 33 meV with 50 Ry cutoff, and within 12 meV with 100 Ry cutoff. Notably, for FF-CD and HL-GPP methods, the convergence rate with respect to dielectric matrix cut-off are similar. The empty circles in Fig. S1 are the extrapolated values of FF-CD calculations based on the convergence curve of HL-GPP results.

To check whether our results and conclusions are affected by different starting points for GW+BSE calculations, we carried out calculations using PBE+U (U=4.0 eV) calculation results as starting wave functions. The GW-BSE calculation results calculated with different starting-point wave functions are compared in Table S1. Although PBE+U predicts a larger Kohn-Sham gap than compared to PBE, the QP band gaps obtained by $G_0W_0$@PBE and $G_0W_0$@PBE+U for $Cs_2TiI_6$ only differ by less than 0.1 eV. In contrast, QP band gap of $Cs_2TiBr_6$ calculated by $G_0W_0$@PBE+U is about 0.5 eV smaller than that calculated with $G_0W_0$@PBE. For both $Cs_2TiI_6$ and $Cs_2TiBr_6$, the energies of lowest-energy dark excitons calculated with $G_0W_0$+BSE@PBE are close to those obtained with $G_0W_0$+BSE@PBE+U. The difference between the results calculated with PBE and PBE+U as starting points is mainly attributed to the differences in the description of dielectric screening effects. In particular, $G_0W_0$@PBE+U predicts a much larger $\varepsilon_\infty$ for $Cs_2TiBr_6$, which also leads to a smaller predicted exciton binding energy of 1.16 eV. Overall, the $G_0W_0$@PBE+U does not alter our main conclusions stated in the main text.

Table S1. QP band gap, optical dielectric constants $\epsilon_\infty$, lowest-energy dark exciton and their binding energy for $Cs_2TiI_6$ and $Cs_2TiBr_6$ with $G_0W_0$@PBE and $G_0W_0$@PBE+U. All the energies are in eV. These results are calculated without including SOC.

|  |  | $Cs_2TiI_6$ |  | $Cs_2TiBr_6$ |  |
| --- | --- | --- | --- | --- | --- |
|  |  | $G_0W_0$@PBE | $G_0W_0$@PBE+U | $G_0W_0$@PBE | $G_0W_0$@PBE+U |
| quasiparticle band gap | Γ - Γ | 1.96 | 1.94 | 3.44 | 2.95 |
|  | X - X | 1.94 | 1.88 | 3.41 | 2.88 |
|  | X - Γ | 1.92 | 1.85 | 3.39 | 2.85 |
| $\varepsilon_\infty$ |  | 4.5 | 4.0 | 3.3 | 5.1 |
| $E_{ex}^{Dark}$ |  | 0.95 | 0.95 | 1.77 | 1.79 |
| $E_b^{Dark}$ |  | 0.88 | 0.99 | 1.58 | 1.16 |

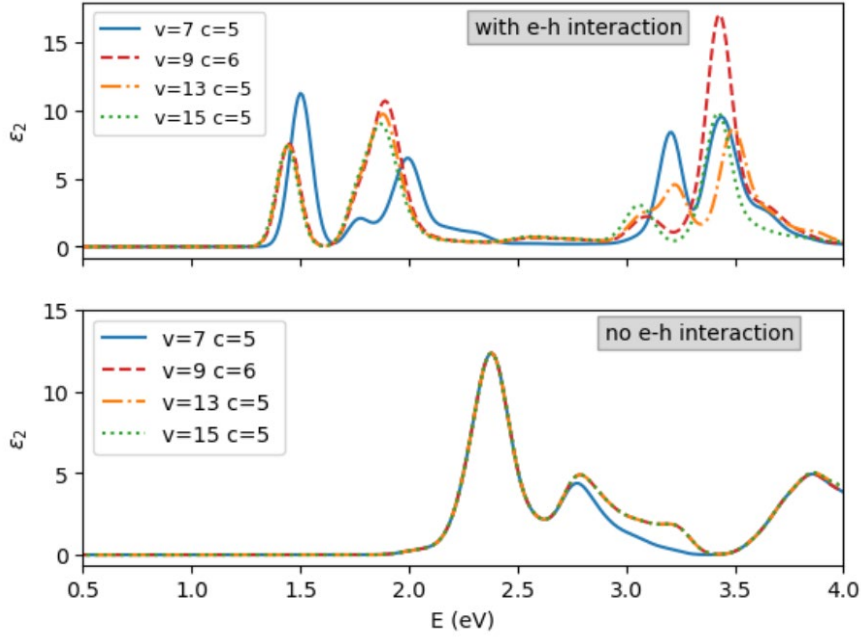

FIG. S2. The optical absorption spectrum for $Cs_2TiI_6$ calculated with different numbers of valence and conduction bands (a) with electron-hole interaction (using GW plus BSE) and (b) without electron-hole interaction (using GW energies). To save computational costs, SOC is not included in this convergence test.

We investigated the convergence of exciton binding energies and frequency-dependent dielectric functions calculated with the Bethe-Salpeter equation. Our tests of convergence are conducted without the spin-orbit coupling. The results show the absorption spectra does not depend sensitively to the density of *k*-point sampling of Brillouin zone. Instead, the number of valence and conduction bands included in the BSE calculations is more important. Fig. S2 shows the optical absorption spectra of $Cs_2TiI_6$ calculated with different numbers of valence and conduction bands include in the BSE calculations. As shown in the upper panel, the low-energy excitonic peaks converge very well with 9 valence bands and 6 conduction bands. For the results calculated without electron-hole interaction, the low-energy optical absorption spectra converge very well with 7 valence bands and 5 conduction bands, as shown in the bottom panel of Fig. S2. Table S2 compares the energies of the lowest-energy and second-lowest spin-singlet excitons calculated different sets of cut-off parameters. The results show that the exciton energies converge very well within 10 meV even if one uses the number of valence bands $N_v = 7$, the number of conduction bands $N_c = 5$, a $4 \times 4 \times 4$ coarse *k*-grid, and a $8 \times 8 \times 8$ fine *k*-grid. The exciton energies also do not depend sensitively on the cut-off energy of dielectric matrix $\varepsilon_{GG'}$. The usage of the coarse and fine *k*-grid in BSE calculations is discussed in the Computational details section.

Table S2. The energies of lowest-energy exciton and second-lowest-energy exciton calculated with different dielectric matrix cut-off, numbers of valence and conduction bands, and different *k*-grid sampling.

| Dielectric matrix cut-off (Ry) | $N_v$ | $N_c$ | Coarse *k*-grid | Fine *k*-grid | Lowest- exciton energy (eV) | Second lowest- exciton energy(eV) |
|---|---|---|---|---|---|---|
| 6 | 7 | 5 | 4×4×4 | 8×8×8 | 1.0516 | 1.1173 |
| | | | | 10×10×10 | 1.0480 | 1.1138 |
| | | | | 13×13×13 | 1.0466 | 1.1126 |
| | 9 | 6 | 4×4×4 | 10×10×10 | 1.0400 | 1.1051 |
| 12 | 7 | 5 | 8×8×8 | 10×10×10 | 1.0427 | 1.1158 |
| | 9 | 6 | 8×8×8 | 10×10×10 | 1.0364 | 1.1088 |
| 18 | 7 | 5 | 4×4×4 | 8×8×8 | 1.0488 | 1.1236 |
| | | | | 10×10×10 | 1.0452 | 1.1201 |
| | | | 6×6×6 | 10×10×10 | 1.0432 | 1.1186 |
| | 9 | 6 | 4×4×4 | 10×10×10 | 1.0372 | 1.1116 |
| | 9 | 8 | 4×4×4 | 10×10×10 | 1.0371 | 1.1114 |
| | 9 | 9 | 4×4×4 | 10×10×10 | 1.0367 | 1.1107 |
| | | | | 13×13×13 | 1.0438 | 1.1189 |
| 25 | 7 | 5 | 4×4×4 | 8×8×8 | 1.0488 | 1.1246 |
| | | | | 10×10×10 | 1.0450 | 1.1212 |
| | | | | 13×13×13 | 1.0437 | 1.1199 |

**Symmetry analysis on dipole-allowed optical transitions within the single-particle picture**

For $\Gamma$ point, the dipole matrix element $\boldsymbol{p} = [p_x, p_y, p_z]$ transforms as the basis of the $\Gamma_{15}$ representation [5]. The decompositions of the direct products of selected double group irreducible representations at $\Gamma$ point are listed as follows

$$\Gamma_8^- \otimes \Gamma_8^+ = \Gamma_1' \oplus \Gamma_2' \oplus \Gamma_{12}' \oplus 2\Gamma_{15} \oplus 2\Gamma_{25}$$

$$\Gamma_8^- \otimes \Gamma_7^+ = \Gamma_{12}' \oplus \Gamma_{15} \oplus \Gamma_{25}$$

$$\Gamma_8^+ \otimes \Gamma_8^+ = \Gamma_1 \oplus \Gamma_2 \oplus \Gamma_{12} \oplus 2\Gamma_{15}' \oplus 2\Gamma_{25}'$$

$$\Gamma_8^+ \otimes \Gamma_7^+ = \Gamma_{12} \oplus \Gamma_{15}' \oplus \Gamma_{25}'$$

$$\Gamma_6^+ \otimes \Gamma_8^+ = \Gamma_{12} \oplus \Gamma_{15}' \oplus \Gamma_{25}'$$

$$\Gamma_6^+ \otimes \Gamma_7^+ = \Gamma_2 \oplus \Gamma_{25}'$$

$$\Gamma_6^- \otimes \Gamma_8^+ = \Gamma_{12}' \oplus \Gamma_{15} \oplus \Gamma_{25}$$

$$\Gamma_6^- \otimes \Gamma_7^+ = \Gamma_2' \oplus \Gamma_{25}$$

Among these decompositions, only $\Gamma_6^- \otimes \Gamma_8^+$, $\Gamma_8^- \otimes \Gamma_7^+$, and $\Gamma_6^- \otimes \Gamma_8^+$ include the $\Gamma_{15}$ representation. Therefore $\Gamma_6^- \to \Gamma_8^+$, $\Gamma_8^- \to \Gamma_7^+$, and $\Gamma_6^- \to \Gamma_8^+$ transitions are dipole allowed, as illustrated in the Fig. 1(d) in the main text.

**Computational details**

**Ground-state calculations**

A first-principles method based on plane-wave basis and the density functional theory (DFT) calculations [6] are employed with the QUANTUM ESPRESSO software package [7]. The Perdew-Burke-Ernzerhof (PBE) exchange-correlation functional [8] is used to account for the exchange-correlation effects. The optimized norm-conserving Vanderbilt pseudopotentials with fully relativistic effects are used for calculations with and without spin-orbit coupling [9–11], respectively, with the following valence electrons configuration: Cs ($5s^2 5p^6 6s^1$), Ti ($3s^2 3p^6 4s^2 3d^2$), Zr ($4s^2 4p^6 5s^2 4d^2$), I ($5s^2 5p^5$) and Br ($4s^2 4p^5$). The Kohn-Sham orbitals are expanded in plane-wave basis sets with a cutoff energy of 80 Ry and the Brillouin zone is sampled using $5 \times 5 \times 5$ uniform *k*-grid.

To check the dependence of results and conclusions on the starting point of GW+BSE calculations, we performed PBE+U (U=4.0 eV) calculations using PARATEC. For PBE+U calculations [12,13], we used Troullier-Martin pseudopotentials [9]. Semi-core electrons are explicitly treated as valence electrons in pseudopotentials for transition metal elements. An energy cut-off of 150 Ry is used for wave functions.

**GW quasiparticle calculations**

The quasiparticle calculations are carried out with the one-shot GW approximation ($G_0W_0$ approximation) implemented in the BERKELEYGW package [14]. Kohn-Sham wave functions are chosen as the starting point for GW plus Bethe-Salpeter equation calculations. We compared the GW calculation results calculated with Hybertsen-Louie generalized plasmon-pole model (HL-GPP) [15] and full-frequency with Contour deformation approach (FF-CD) [16,17]. Our results show HL-GPP predicts a quasiparticle band gap that is ~ 0.3 eV smaller than that predicted with FF-CD method. The results presented in the main text are obtained with FF-CD method. We computed the dielectric matrices on a 4 × 4 × 4 $k$-point mesh with a polarizability cutoff of 45 Ry. We used 1200 empty states in combination with the static-remainder approach [18] to calculate the quasiparticle energies. The QP band gaps are converged to within 33 meV with these settings.

For G$_0$W$_0$@PBE+U calculations, we used a modified version of BERKELEYGW code, which implements the energy-integration method [4] to include all the empty states in GW calculations.

**Optical properties**

The calculations of optical properties are carried out by solving the Bethe-Salpeter equation (BSE) within the Tamm-Dancoff approximation [19,20], as implemented in the BERKELEYGW package [14]. The optical excitations are calculated by direct diagonalizing BSE:

$$\left(E_{ck}^{QP} - E_{vk}^{QP}\right)A_{vck}^S + \sum_{v'c'k'}\langle vck|K^{eh}|v'c'k'\rangle = E_{ex}^S A_{vck}^S$$

where $E_{nk}^{QP}$ are the quasiparticle energies calculated with G$_0$W$_0$ approximation, $A_{vck}^S$ is the eigenvectors of excitons (in the quasiparticle state representation), $E_{ex}^S$ is the exciton energy, and $K^{eh}$ is the electron–hole interaction kernel. To save computational cost while without sacrificing the accuracy, the quasiparticle energies are approximated by using scissor shifts. This is a good approximation since the frontier conduction (valence) bands have similar characters of atomic orbitals.

Here, to understand the role of spin-orbit coupling, we solved BSE for the case with and without considering spin-orbit coupling effects. In the case when spin-orbit coupling effects are neglected, spin-triplet and spin-singlet excitations are calculated separately by diagonalizing BSE by setting $K_{singlet}^{eh} = K_x + K_d$ and $K_{triplet}^{eh} = K_d$, where $K_x$ and $K_d$ are exchange and direct interaction terms, respectively [20]. The matrix elements of the BSE Hamiltonian are explicitly calculated using 9 valence and 9 conduction bands on a coarse 4 × 4 × 4 $k$-grid and subsequently interpolated onto a finer 10×10×10 $k$-grid using 7 valence and 5 conduction bands, which is then diagonalized to yield the exciton states and the resulting absorption spectrum. We find the exciton binding energies as well as the frequency-dependent dielectric functions converge very fast with the density of $k$-grid, because the exciton wave functions are highly extended (evenly distributed) in the reciprocal space.

**Density functional perturbation theory calculations**

We calculate the phonon dispersion spectra and the LO phonon frequency within density functional perturbation theory (DFPT) [21], as implemented in the QUANTUM ESPRESSO package with Phonopy [22]. The phonon dispersion spectra are obtained with non-analytical term correction, as shown in Fig. S7. And the ionic contributions to the frequency dependent dielectric function reported in the Table 1 of the main manuscript are carried out with the Vienna ab initio simulation package (VASP) [23]. SOC are not considered in DFPT calculations.

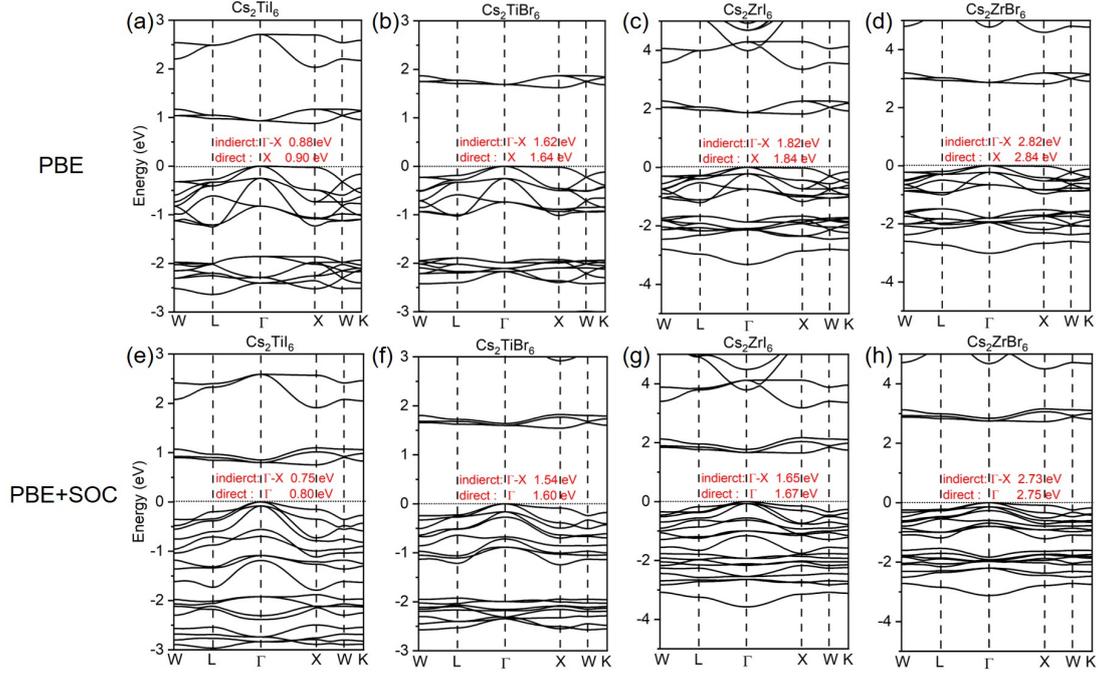

FIG. S3. The DFT band structure along W (0.5, 0.25, 0.75) - L (0.5, 0.5, 0.5) - Γ (0, 0, 0) - X (0.5, 0.0, 0.5) - W (0.5, 0.25, 0.75) – K (0.375, 0.375, 0.75) of four VODPs within (a-d) PBE functional and (e-h) PBE+SOC functional.

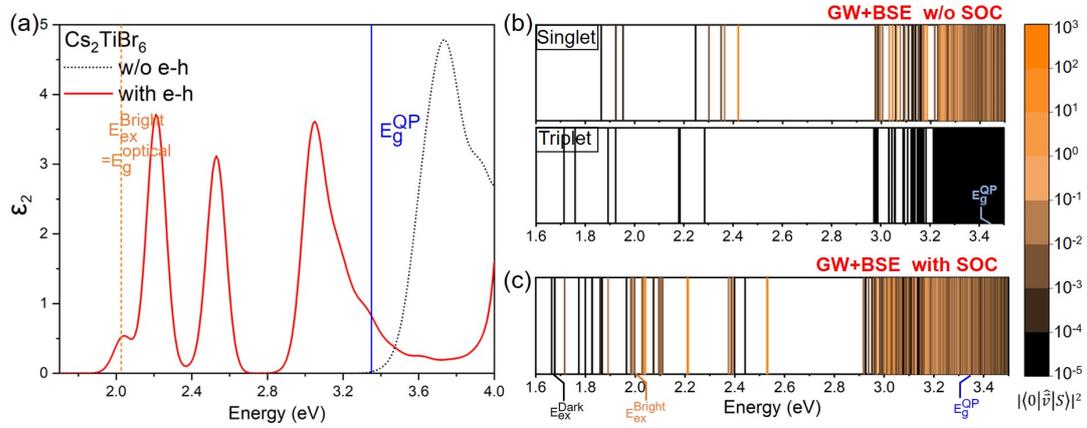

FIG. S4. (a) Optical absorption spectrum of $Cs_2TiBr_6$, calculated without electron−hole interaction (black dashed line) and with electron-hole interaction (red solid line). A gaussian smearing of 0.05 eV is used for calculating the spectra. The lowest-energy bright exciton and fundamental band gap are marked in orange and blue lines, respectively. (b) Energies of singlet and triplet excitons calculated with a GW+BSE approach without SOC effects. (c) Energies of excitons calculated with a GW+BSE approach including SOC effects.

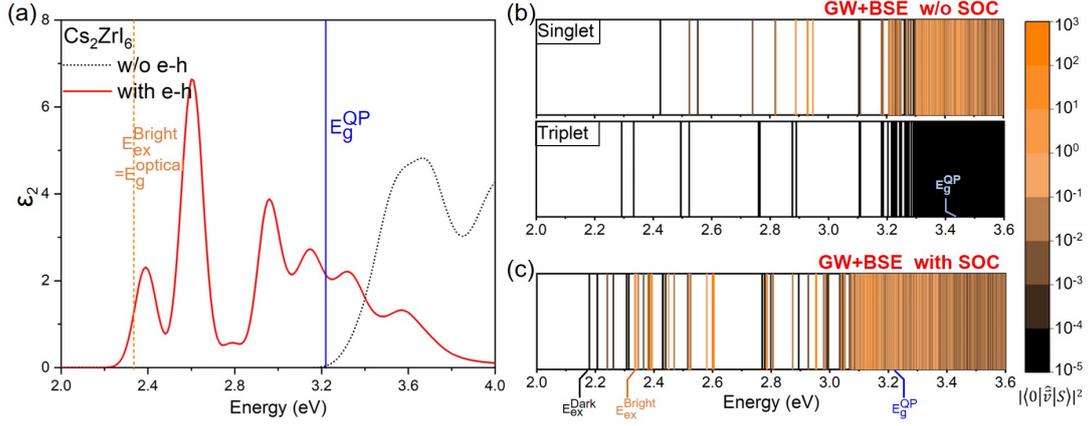

FIG S5. (a) Imaginary part of frequency-dependent dielectric function of $Cs_2ZrI_6$, calculated without electron−hole interaction (black dashed line) and with electron-hole interaction (red solid line). A gaussian smearing of 0.05 eV is used for calculating the spectra. The lowest-energy bright exciton and fundamental band gap are marked in orange and blue lines, respectively. (b) Energies of singlet and triplet excitons calculated with a GW+BSE approach without SOC effects. (c) Energies of excitons calculated with a GW+BSE approach including SOC effects.

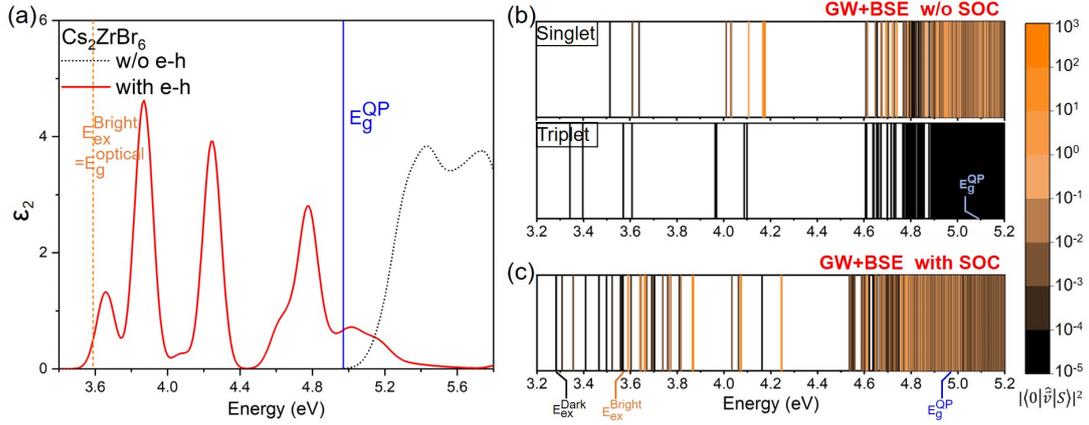

FIG. S6. (a) Optical absorption spectrum of $Cs_2ZrBr_6$, calculated without electron−hole interaction (black dashed line) and with electron-hole interaction (red solid line). A gaussian smearing of 0.05 eV is used for calculating the spectra. The lowest-energy bright exciton and fundamental band gap are marked in orange and blue lines, respectively. (b) Energies of singlet and triplet excitons calculated with a GW+BSE approach without SOC effects. (c) Energies of excitons calculated with a GW+BSE approach including SOC effects.

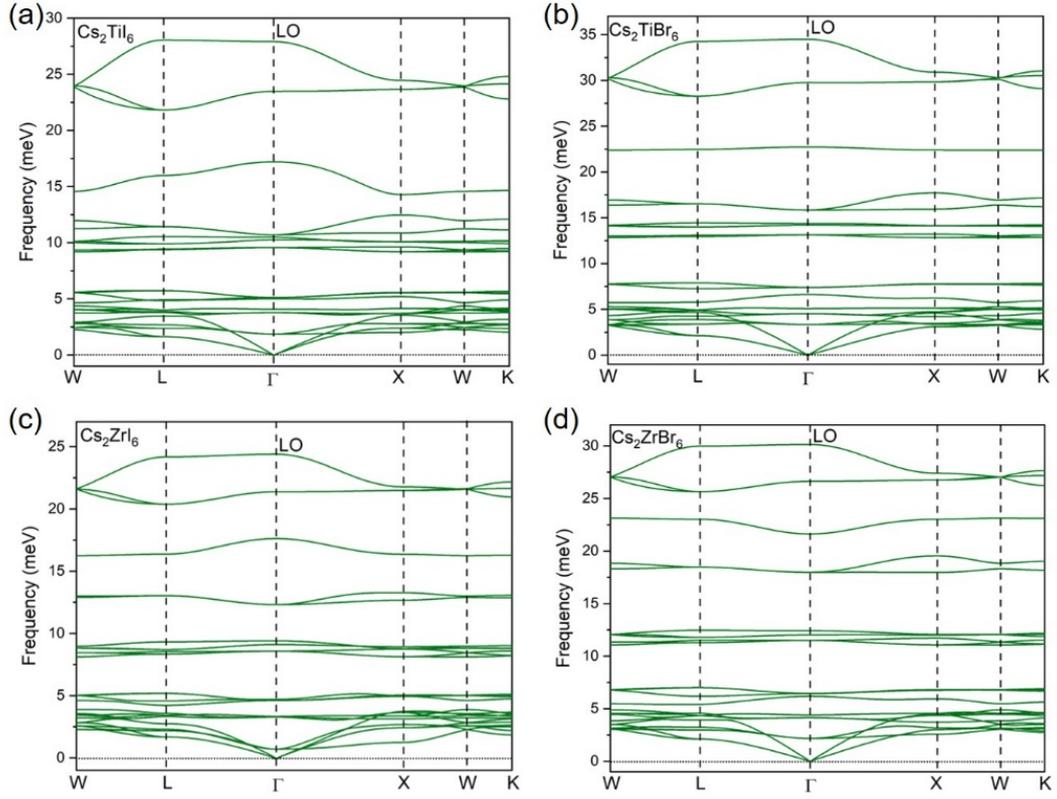

FIG. S7. The phonon dispersion spectra with non-analytical term correction of four VODPs. longitudinal-optical (LO) phonon is marked in the figure and W (0.5, 0.25, 0.75), L (0.5, 0.5, 0.5), Γ (0, 0, 0), X (0.5, 0.0, 0.5) and K (0.375, 0.375, 0.75) are high-symmetry $k$-points in the first Brillouin zone.